\documentclass[conference]{IEEEtran}
\IEEEoverridecommandlockouts
\usepackage{cite}
\usepackage{url}
\usepackage{amsmath,amssymb,amsfonts}
\usepackage{algorithmic}
\usepackage{graphicx}
\usepackage{textcomp}
\usepackage{xcolor}
\usepackage{algorithm}
\usepackage{algorithmic}
\usepackage{subfigure}
\def\BibTeX{{\rm B\kern-.05em{\sc i\kern-.025em b}\kern-.08em
    T\kern-.1667em\lower.7ex\hbox{E}\kern-.125emX}}

\renewcommand{\figurename}{Fig.}
\renewcommand{\tablename}{Table}
\newcommand{\figref}[1]{Fig.\hspace{1mm}\ref{#1}}
\newcommand{\tabref}[1]{Table\hspace{1mm}\ref{#1}}

\newcommand{\algref}[1]{Algorithm.\hspace{1mm}\ref{#1}}

\begin{document}

\title{Efficient feature embedding of 3D brain MRI images for content-based image retrieval\\with deep metric learning}
\author{
\IEEEauthorblockN{
  Yuto Onga\IEEEauthorrefmark{1}, 
  Shingo Fujiyama\IEEEauthorrefmark{1},
  Hayato Arai\IEEEauthorrefmark{1},
  Yusuke Chayama\IEEEauthorrefmark{1},
  Hitoshi Iyatomi\IEEEauthorrefmark{1},
  Kenichi Oishi\IEEEauthorrefmark{2}
}
\IEEEauthorblockA{
  Email: \{yuto.onga.3u, shingo.fujiyama.5a, hayato.arai.5t, yusuke.chayama.2t\}@stu.hosei.ac.jp,\\iyatomi@hosei.ac.jp, koishi2@jhmi.edu}
  \IEEEauthorrefmark{1}Applied Informatics, Graduate School of Science and Engineering, Hosei University, Tokyo, Japan\\
  \IEEEauthorrefmark{2}Johns Hopkins University School of Medicine, MD, USA\\
}

\maketitle
\makeatletter

\def\ps@IEEEtitlepagestyle{%
  \def\@oddfoot{\mycopyrightnotice}%
  \def\@evenfoot{}%
}
\def\mycopyrightnotice{%
  {\footnotesize 978-1-7281-0858-2/19/\$31.00 © 2019 IEEE\hfill}
  \gdef\mycopyrightnotice{}
}

\begin{abstract}
  Increasing numbers of MRI brain scans, improvements in image resolution, and advancements in MRI acquisition technology are causing significant increases in the demand for and burden on radiologists' efforts in terms of reading and interpreting brain MRIs. Content-based image retrieval (CBIR) is an emerging technology for reducing this burden by supporting the reading of medical images. High dimensionality is a major challenge in developing a CBIR system that is applicable for 3D brain MRIs. In this study, we propose a system called disease-oriented data concentration with metric learning (DDCML). In DDCML, we introduce deep metric learning to a 3D convolutional autoencoder (CAE). Our proposed DDCML scheme achieves a high dimensional compression rate (4096:1) while preserving the disease-related anatomical features that are important for medical image classification. The low-dimensional representation obtained by DDCML improved the clustering performance by 29.1\% compared to plain 3D-CAE in terms of discriminating Alzheimer's disease patients from healthy subjects, and successfully reproduced the relationships of the severity of disease categories that were not included in the training.

\end{abstract}


\section{Introduction}

  \begin{figure*}[t]
    \begin{center}
    \includegraphics[width=0.95\linewidth]{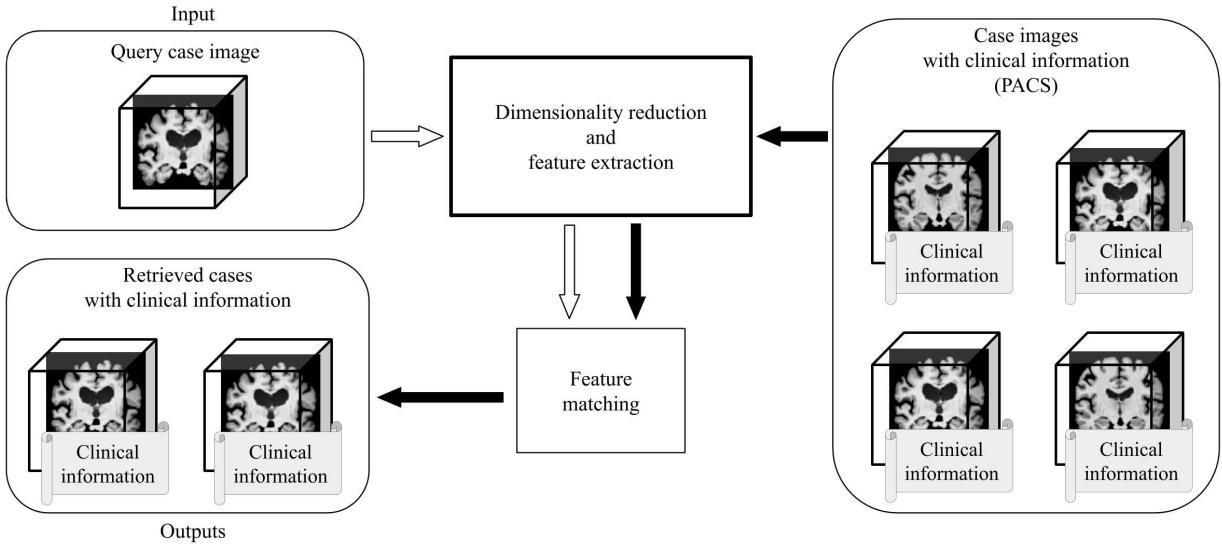}
    \caption{An overview of content-based image retrieval system in brain images. In this study, we propose new dimensionality reduction technique, which is an operation surrounded by a thick frame.}
    \label{fig:CBIR}
    \end{center}
  \end{figure*}

  A brain MRI provides information about in vivo brain anatomy and pathology, which are essential in the diagnosis and treatment of neurological diseases. The numbers of MRI scanners are increasing worldwide, and image quality is improving in terms of resolution and contrast, with technological advancements being made in scanners, scan sequences and parameters. However, these  improvements in both the quantity and quality of brain MRI acquisition are increasing the burden on the radiologists who read and interpret these images. An image reading support system is therefore highly desirable to reduce the burden on radiologists and to improve the quality of medical practice.
  Currently, a very large number of brain MRIs are being stored as digital data in picture archiving and communication systems (PACSs), with their corresponding clinical information. A PACS enables effective data sharing of MRIs and communication among physicians and patients, thus providing the information necessary to make medical decisions for each patient. The natural extension of the role of a PACS is to use the big data collected through medical practice as a resource to support the reading of and decision making on brain MRIs. This system would enable learning from past medical decisions to maximize the quality of current health care.
  Content-based image retrieval (CBIR) is regarded as one of the most promising technologies for utilizing the big data related to brain MRIs stored in PACS in medical practice. The ideal CBIR would allow users to submit their patients' brain MRIs, to search the images stored on the PACS server and to retrieve MRIs with pathological features that are similar to those in the submitted image. This system would provide clues for diagnosis and prognosis by creating a list of potential diagnoses and their probabilities, and by reporting prognostic information obtained from the corresponding electronic medical records. \figref{fig:CBIR} shows an overview of ideal CBIR system that has not been achieved yet. To handle the big data involved, a machine learning framework needs to be introduced. However, there are several issues that hinder the application of machine learning to CBIR for clinical brain MRIs.
  One of the major issues is the high dimensionality. A high-resolution 3D anatomical brain MRI is becoming a common modality for evaluating neurodegenerative diseases such as Alzheimer's disease (AD), and each image typically contains more than a million pixels per scan. This rich anatomical and pathological information creates a problem known as “the curse of dimensionality". Attempts to apply machine learning approaches such as clustering, regression, and classification to high-dimensional raw data will fail, since overfitting will occur. Thus, the number of features used for CBIR must be reduced while preserving the disease-related pathological features that are important for medical image classification. 
  Several pioneering efforts have been made to construct CBIR systems that are applicable to brain MRIs, including schemes such as a region-specific bag of visual words~\cite{huang2012retrieval}, KD tree or KNN~\cite{arakeri2013intelligent, srilakshmi2015performance}, a Gabor local mesh pattern~\cite{murala2013local}, and singular value decomposition (SVD)~\cite{lyra2012improved}. While these techniques have shown attractive results, their scope is limited (for example, they are applicable to only limited types of diseases~\cite{srilakshmi2015performance, murala2013local}, or the depth information cannot be taken into account since slices were used as input~\cite{huang2012retrieval, arakeri2013intelligent, lyra2012improved}), since they are based on traditional machine learning techniques with manual feature engineering. Hence, these technologies are not suitable for handling the images stored in a PACS, which may contain a wide variety of diseases with different pathological features. An atlas-based brain MRI parcellation approach, in which the anatomical and pathological features of the brain are extracted from local brain volumes or intensities obtained from approximately 250 anatomical structures, has demonstrated excellent performance in terms of retrieval when applied to neurodegenerative diseases such as primary progressive aphasia~\cite{faria2015content, qin2013gross}, AD, Huntington's disease, and spinocerebellar ataxia~\cite{qin2013gross}. The major advantage of the atlas-based approach is the anatomically meaningful and highly effective dimension reduction, which makes the biological and pathological interpretation of the CBIR results straightforward. However, the generalizability of this approach to other neurological diseases has yet to be investigated. 
  Recent advancements in the field of computer vision, and particularly in convolutional neural networks (CNN), have allowed the fully automated extraction of the image features necessary for classification in the learning phase. Several automated diagnosis techniques using 3D brain MRI images have been proposed for Alzheimer's and related diseases; these are based on the CNN framework, which has shown reasonable results~\cite{korolev2017residual, esmaeilzadeh2018end}. 
  We have previously proposed a dimension reduction technique for high-resolution 3D brain MRIs using a 3D convolutional autoencoder (3D-CAE), and have achieved a high dimension reduction rate, compressing approximately five million dimensional inputs to only 150 dimensions while preserving the clinically relevant neuroradiological features~\cite{arai2018significant}. However, the evaluation of this approach was based on qualitative observations of the reconstructed images performed by a neurologist. In follow-up experiments, we applied the 3D-CAE to CBIR, and became aware that this low-dimensional representation was affected by normal anatomical variations such as brain gyrification patterns, rather than disease-related pathological features. Since the goal of our clinical CBIR system is to search for and retrieve brain MRIs based on their pathological similarities, we needed to modify the 3D-CAE method to focus on extracting pathological features while ignoring brain features that are not related to disease pathology.
  In this paper, we introduce metric learning to overcome the current limitations of 3D-CAE. The basic concept of metric learning is that data with similar properties (i.e. the same disease) in real space should be located near to each other in the low-dimensional space. Metric learning has been successful in a wide range of applications including search technology~\cite{sohn2016improved, oh2016deep}. Song et al. demonstrated the capability of this approach in clustering data belonging to categories that were not included in the training data~\cite{oh2016deep}. This feature is advantageous in finding similar MRIs of diseases that were not included in the training dataset. Hoffer et al. reported that the application of semi-supervised learning with unannotated data to metric learning improved the accuracy of classification results~\cite{hoffer2016semi}. This feature is important for medical image classification, in which the amount of training data with professional annotations is limited. We hypothesize that our novel method that incorporates metric learning into the 3D-CAE, called disease-oriented data concentration with metric learning (DDCML), provides a clinically meaningful low-dimensional representation of a brain MRI while preserving disease-related pathological features.
  The main contribution of this paper is to provide a practical method for the low-dimensional representation of 3D brain MRI images for clinical CBIR. By using only data from AD and healthy (cognitively normal, CN) patients for training, the proposed DDCML provides a low-dimensional representation that is preferable for CBIR, which not only separates AD and NC by more than 80\% using a simple K-means algorithm but also provides an appropriate distribution of untrained medical conditions according to their severity, such as early and late mild cognitive impairment (EMCI/LMCI) and subjective memory concerns (SMC).

\section{Material and pre-processing}
  We used the Alzheimer's Disease Neuroimaging Initiative-2 (ADNI2) dataset in this experiment, which was created for the purpose of early detection, treatment, and research to study AD. Each image contains $256 \times 256 \times 170$ pixels and falls into one of the following classes: patients with AD, EMCI, LMCI or SMC, and healthy patients (CN). 
  AD is one of the major types of dementia; however, SMC is a subjective symptom of memory decline, and no medical symptoms of dementia are observed in this condition. From a medical perspective, the progression of dementia can be represented as CN$\fallingdotseq$SMC$\leq$EMCI$\leq$LMCI$\leq$AD.
  We performed skull removal and volume correction as a pre-processing stage using MRICloud~\cite{mori2016mricloud}\footnote{\url{https://www.mricloud.org}}. The size of each resulting preprocessed image was $181 \times 217 \times 181$ pixels. An additional preprocessing step was performed to obtain the optimal shape for passing to our NN model. Based on previous findings, downsampling was performed, and we also removed the margins to obtain images with a final input size of $80 \times 96 \times 80$ pixels similar to other studies analyzing 3D brain MRI images~\cite{korolev2017residual}.
  MRICloud occasionally failed to perform skull removal, and we excluded those cases from our experiments via visual assessment by a physician. The final numbers of images in our dataset were 674, 1,121, 147, 280 and 33 for AD, NC, EMCI, LMCI, and SMC, respectively, resulting in a total of 2,555 images.

  \begin{figure*}[]
    \includegraphics[width=0.99\linewidth]{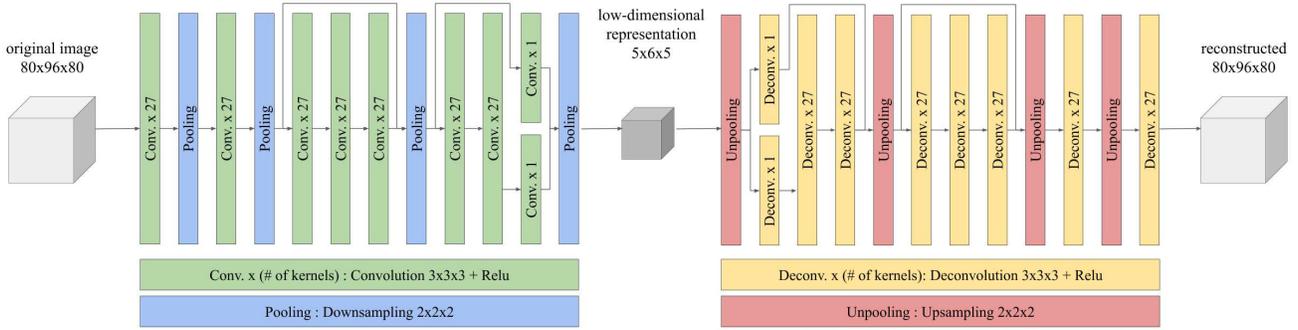}
    \caption{Our proposed CAE network architecture.}
    \label{fig:network_architecture}
  \end{figure*}

\section{DDCML: The proposed CBIR model}
  Our proposed DDCML method involves intensity normalization and efficient dimensional reduction using 3D-CAE with metric learning.

  \subsection{Intensity normalization}
  \label{subsec:I-norm}
  The brightness and contrast of the MRI images are determined by the radiation technician or radiologist at the time of storage. The range of intensity was very diverse for the dataset used. Our preliminary experiments indicated that this variance reduced the performance. We therefore standardized the intensity of the brain area in each case, so that each image had a target average intensity $\mu$ and tolerance $\epsilon$, using iterative gamma correction. \algref{alg:gamma_correction} shows the details of this process. 

    \begin{algorithm}                      
      \caption{Make average image intensity $x$ close to $\mu$ with tolerance $\epsilon$}         
      \label{alg:gamma_correction}                          
      \begin{algorithmic}                  
        \WHILE{$\|x-\mu\| \leq \epsilon$}
        \STATE $\gamma \leftarrow \frac{\mu}{\bar{x}}$
        \STATE $x \leftarrow 255 \left(\frac{x}{255}\right)^{\left(\frac{1}{\gamma}\right)}$
        \ENDWHILE
      \end{algorithmic}
    \end{algorithm}

  \subsection{3D-CAE with metric learning}
  An autoencoder is a technique that maps high-dimensional input data to a low-dimensional representation. It has a symmetrical structure consisting of an encoder and a decoder: the former encodes an input $\mathbf{x} \in R^D$ to a low-dimensional representation $\mathbf{z} \in R^{D_\mathbf{z}} (D \gg D_\mathbf{z})$, while the latter decodes from z to the output $\hat{\mathbf{x}}$ with the same number of dimensions as the input. 
A typical autoencoder consists of neural networks and is trained to have the same inputs and outputs; a traditional backpropagation algorithm can therefore be applied, and training does not require an external training signal.

The function of the encoder can be expressed formally as 

    \begin{equation}
      \mathbf{z}=f(\mathbf{W} \mathbf{x}+\mathbf{b}),
    \end{equation}

    where $\mathbf{W}$ and $b$ are learnable parameters and $f$ is a trainable nonlinear function. The decoder performs inverse mapping of the encoder as 

    \begin{equation}
      \hat{\mathbf{x}}=\tilde{f}(\tilde{\mathbf{W}} \mathbf{z}+\tilde{\mathbf{b}}),
    \end{equation}

    where  $\hat{\mathbf{x}}$ is an output, and $\tilde{\mathbf{W}}$, $\tilde{b}$ and $\tilde{f}$ are the learnable parameters and function of the decoder as before. Common autoencoders include a root mean squared error (RMSE) between the input and output as a loss function, 

    \begin{equation}
      L_{RMSE}(\mathbf{W}) =\frac{1}{D} \sum_{d=1}^{D}\left\|\mathbf{x}_{d}-\hat{\mathbf{x}}_{d}\right\|^{2}.
    \end{equation}

    As a result, the autoencoder acquires a low-dimensional representation of the input while retaining the data characteristics. CAE incorporates the CNN mechanism, which has a proven track record in the field of computer vision. 3D-CAE is a 3D-enhanced version of CAE that is suitable for handling large volumes of data as it can directly process MRI images of the brain as 3D images rather than continuous 2D slices.

    In order to construct a CBIR system, the distance between the features of a given disease in the low-dimensional feature space should be small. We therefore introduced the idea of deep metric learning into our 3D-CAE; more specifically, we added constraints to reduce the Euclidean distance between features with the same label and increase the distance between features with different labels. Following~\cite{hoffer2016semi}, we calculate the embedded similarity between the input data $\mathbf{x}$ and data from each class $\mathbf{x}_i (i \in {1\ldots c}$; there are $c$ classes in total) in the low-dimensional feature space:

    \begin{equation}
      P \left( \mathbf{x} ; \mathbf{x} _ { 1 } , \ldots , \mathbf{x} _ { c } \right) _ { i } = \frac { e ^ { - \left\| f ( \mathbf{x} ) - f \left( \mathbf{x} _ { i } \right) \right\| ^ { 2 } } } { \sum _ { j = 1 } ^ { c } e ^ { - \left\| f ( \mathbf{x} ) - f \left( \mathbf{x} _ { j } \right) \right\| ^ { 2 } } } , i \in \{ 1 \ldots c \}
    \end{equation}

    Note here that each data point $\mathbf{x}_1, \mathbf{x}_2, \ldots, \mathbf{x}_c$ is randomly selected from each associated class. $P(\mathbf{x}; \mathbf{x}_1, \ldots, \mathbf{x}_c)_i$ represents the probability of a data point $\mathbf{x}$ being classified into class $i$. Here, $f$ indicates the operation of the encoder. By taking the cross-entropy of the one-shot representation $I\in R^c$ of $\mathbf{x}$ and $P$, the "discriminative loss" $L_{Disc}$ is obtained:

    \begin{equation}
      L_{Disc} \left( \mathbf{x} , \mathbf{x} _ { 1 } , \ldots , \mathbf{x} _ { c } \right) = H \left( I ( \mathbf{x} ) , P \left( \mathbf{x} ; \mathbf{x} _ { 1 } , \ldots , \mathbf{x} _ { c } \right) \right).
    \end{equation}

    The purpose of this loss is to ensure that samples belonging to the same class are mapped closer together in the low-dimensional feature space than samples from different classes. Finally, the total loss function of the proposed 3D-CAE scheme is defined as a weighted sum of these two loss criteria with parameter $\alpha$:

    \begin{equation}
      L = L_{RMSE}(\mathbf{W}) + \alpha L_{Disc} \left( \mathbf{x} , \mathbf{x} _ { 1 } , \ldots , \mathbf{x} _ { c } \right)
    \end{equation}

    In our experiments, we used $\alpha=1$.

\section{Experiments adn Results}

  \subsection{3D-CAE architecture}
  The 3D-CAE architectures used in our experiments are shown in \figref{fig:network_architecture}. Our encoder is composed of four blocks: the upper two blocks consist of a convolution and a pooling layer, while the lower two blocks consist of three convolution layers and one pooling layer. Residual bypasses~\cite{he2016deep} were inserted at two locations: between the output of the second block and the output of the third block, and between the output of the third block and the third convolution layer of the fourth block. The kernel size was fixed at 27 for all convolution layers except in the innermost layer with size of 1, and the innermost layer was obtained by flattening the precedent $5\times6\times5$ output neurons. This is namely $D_{z}$ in our setting is 150, and the dimensional compression ratio is ($80\times80\times96$):150, i.e. 4,096:1. We designed the decoder to be symmetrical with the encoder using deconvolution and unpooling layers. For the sake of comparison, we used this CAE architecture both for the baseline (i.e. plain 3D-CAE) and our proposed DDCML system.

  \subsection{Training and Evaluation}
  In our experiments, the AD and CN cases of the ADNI dataset are used for training, and the remaining LMCI, EMCI, and SMC cases are used to verify whether the trained model works effectively for unknown diseases. In this study, the performance was evaluated using group five-fold cross-validation, which uses different splits in the patient data into training and the validation sets. This is to prevent bias caused by separating similar types of data into the training and evaluation sets. We evaluated the capability of our proposed DDCML scheme from two perspectives. The first was a quantitative measure of the reconstructed image to determine how much information is preserved in the low-dimensional representation $\mathbf{z}$. We evaluated these images using the RMSE and SSIM. The second is the availability of our low-dimensional representation for CBIR tasks. We clustered these with the K-means algorithm and measured how well the generated clusters were divided between the AD and CN cases. We also investigated the data distribution between the unlabeled LMCI, EMCI and SMC cases (i.e. those excluded from the training of 3D-CAE) in our low-dimensional feature space.
  In addition, we visualized our low-dimensional representation using t-SNE~\cite{maaten2008visualizing}, i.e. further compressing our 150 dimensions of data into two dimensions and visually examining the data distribution.

\section{Results}
  \tabref{tab:err_and_accs} gives a performance comparison of the image reconstruction and clustering obtained with the K-means algorithm (K=2) using plain 3D-CAE (i.e. without metric learning) and our DDCML scheme. In each hold, K-means clustering is carried out with 10 different initial seeds, and the scores in the table are the average and standard deviation.

\begin{figure}[]
    \begin{flushright}
    \subfigure[original]{\includegraphics[width=0.32\linewidth]{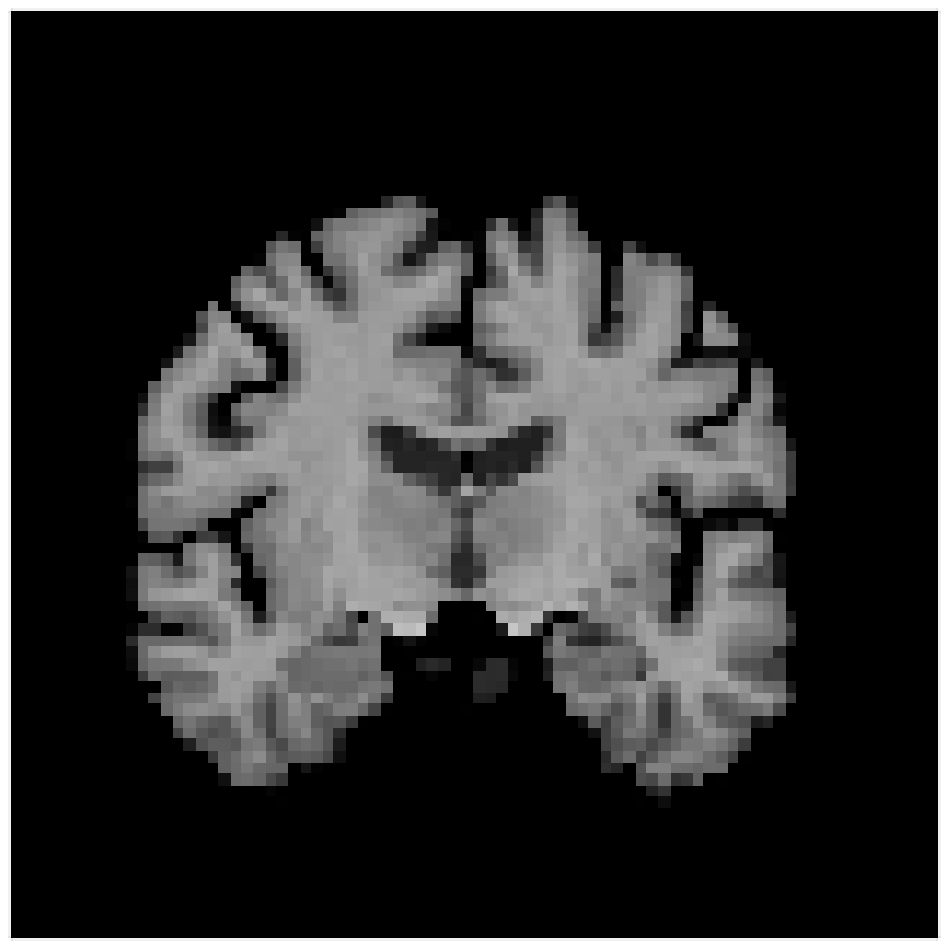}}
    \subfigure[plain CAE]{\includegraphics[width=0.32\linewidth]{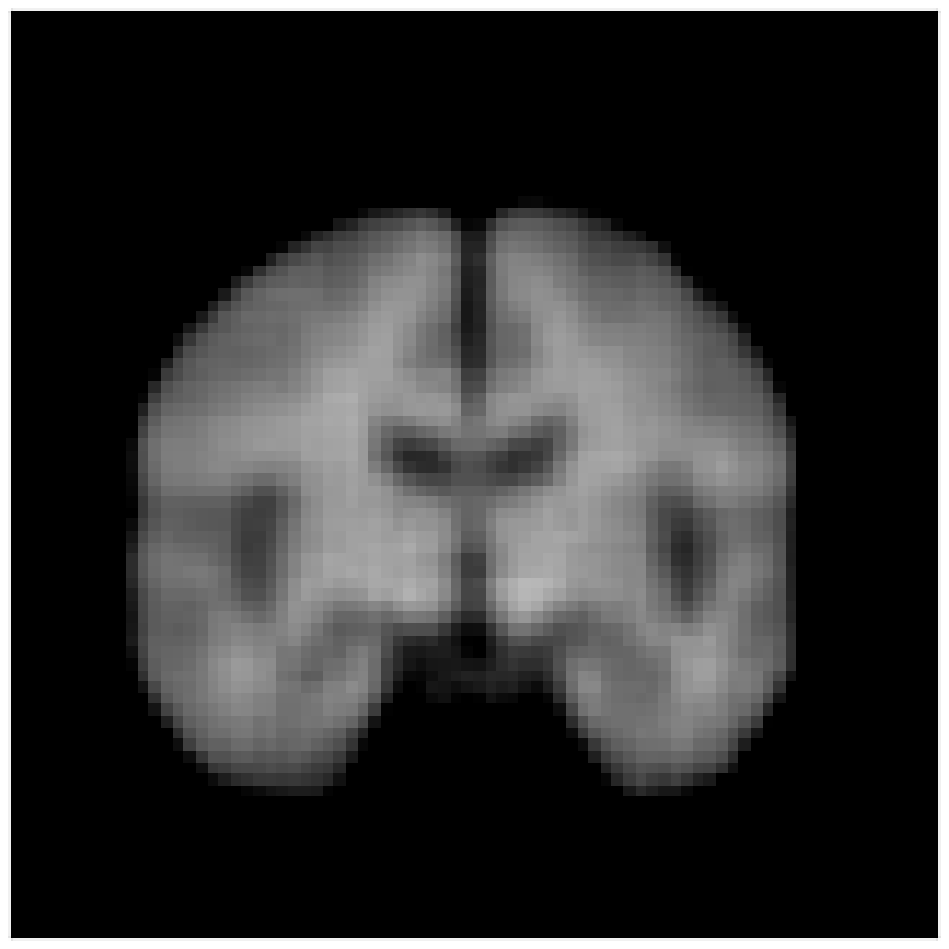}}
    \subfigure[plain CAE + I-norm]{\includegraphics[width=0.32\linewidth]{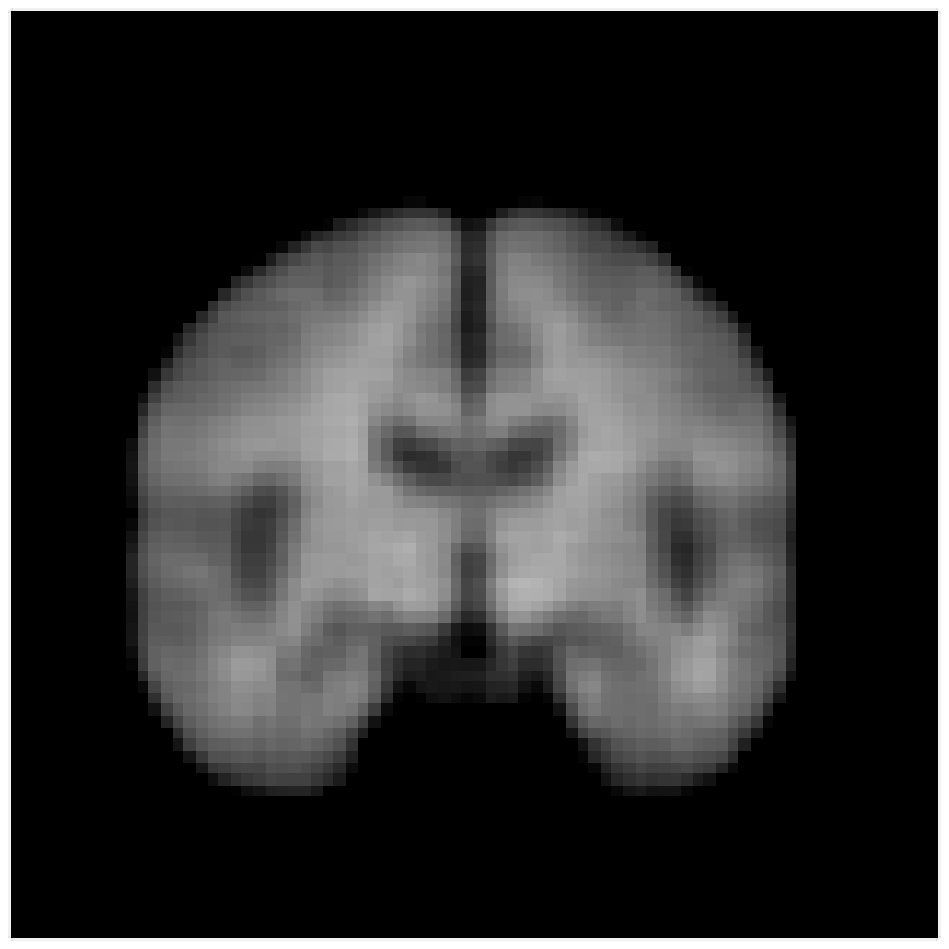}}
    \subfigure[DDCML]{\includegraphics[width=0.32\linewidth]{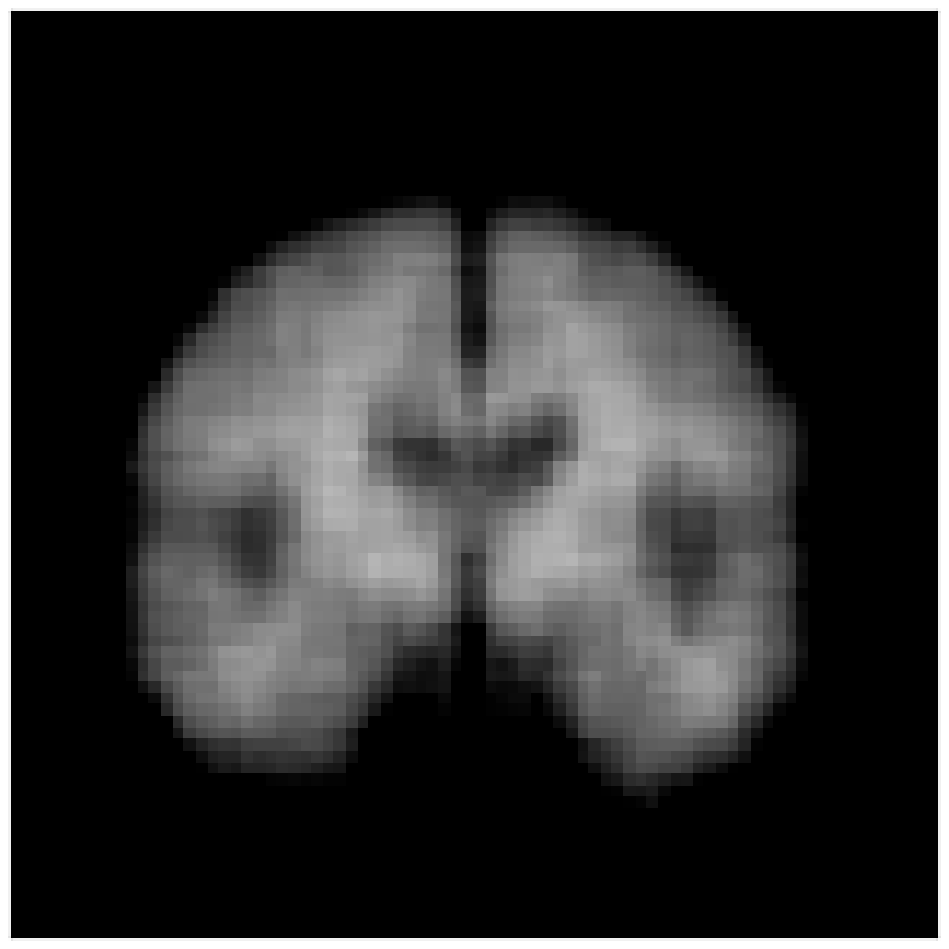}}
    \subfigure[DDCML + I-norm]{\includegraphics[width=0.32\linewidth]{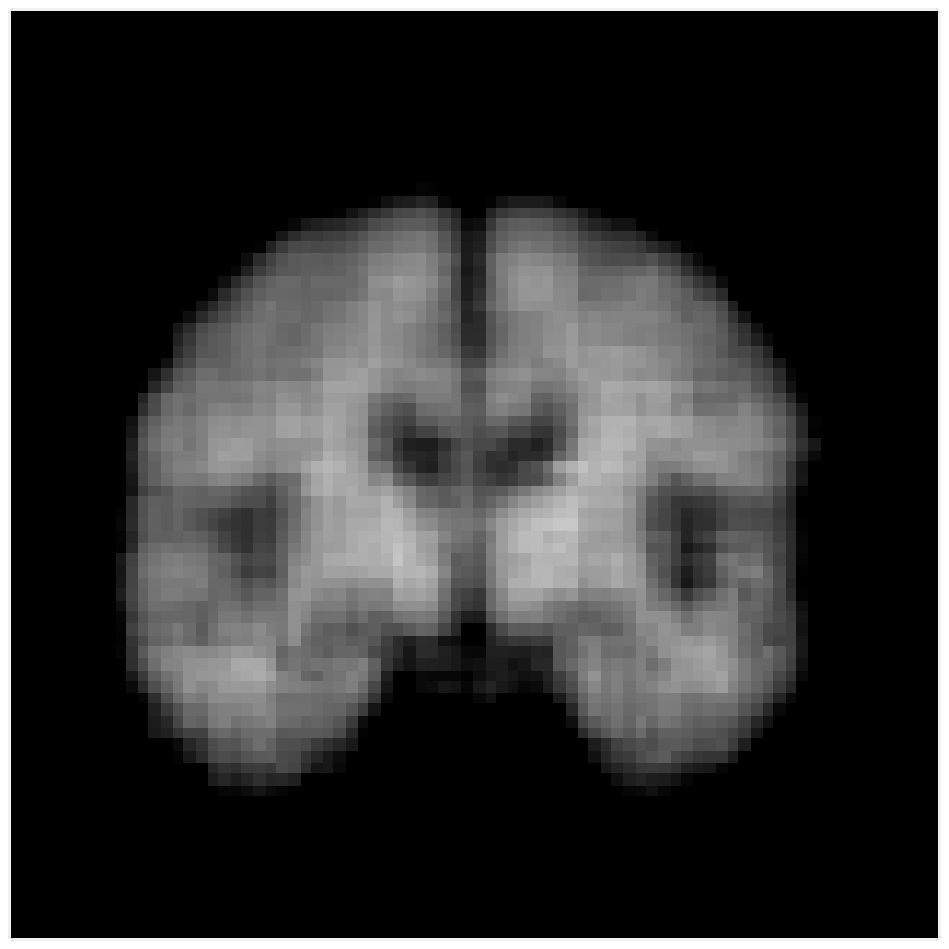}}
    \end{flushright}
    \caption{A slice of the original image (a), slices of the reconstructed images of plain CAE (b, c), slices of the reconstructed images of our DDCML (d, e).}
    \label{fig:reconstructs}
  \end{figure}

  \figref{fig:reconstructs} shows an example of the reconstructed brain images using (b) plain 3D-CAE; (c) 3D-CAE with intensity normalization; (d) 3D-CAE with metric learning; and (e) DDCML (3D-CAE with intensity normalization and metric learning).

  \begin{table}[]
  \begin{center}
  \caption{Comparision of model performance}
  \label{tab:err_and_accs}
  \begin{tabular}{lrrr}
  \hline
  \multicolumn{1}{r}{}                                                               & RMSE(\%)$\downarrow$ & SSIM $\uparrow$ & clustering accuracy(\%) $\uparrow$ \\ \hline
  plain CAE                          & 7.27                 & 0.967           & 52.4($\pm$1.09)                         \\
  plain CAE + I norm & 7.36                 & 0.966           & 55.2($\pm$3.58)                         \\
  DDCML                              & 8.79                 & 0.949           & 80.9($\pm$2.38)                         \\
  DDCML + I norm    & 8.51                 & 0.953           & 81.5($\pm$2.76)                         \\ \hline
  \end{tabular}
  \end{center}
  \end{table}

\begin{figure}[]
    \subfigure[plain CAE]{\includegraphics[width=0.49\linewidth]{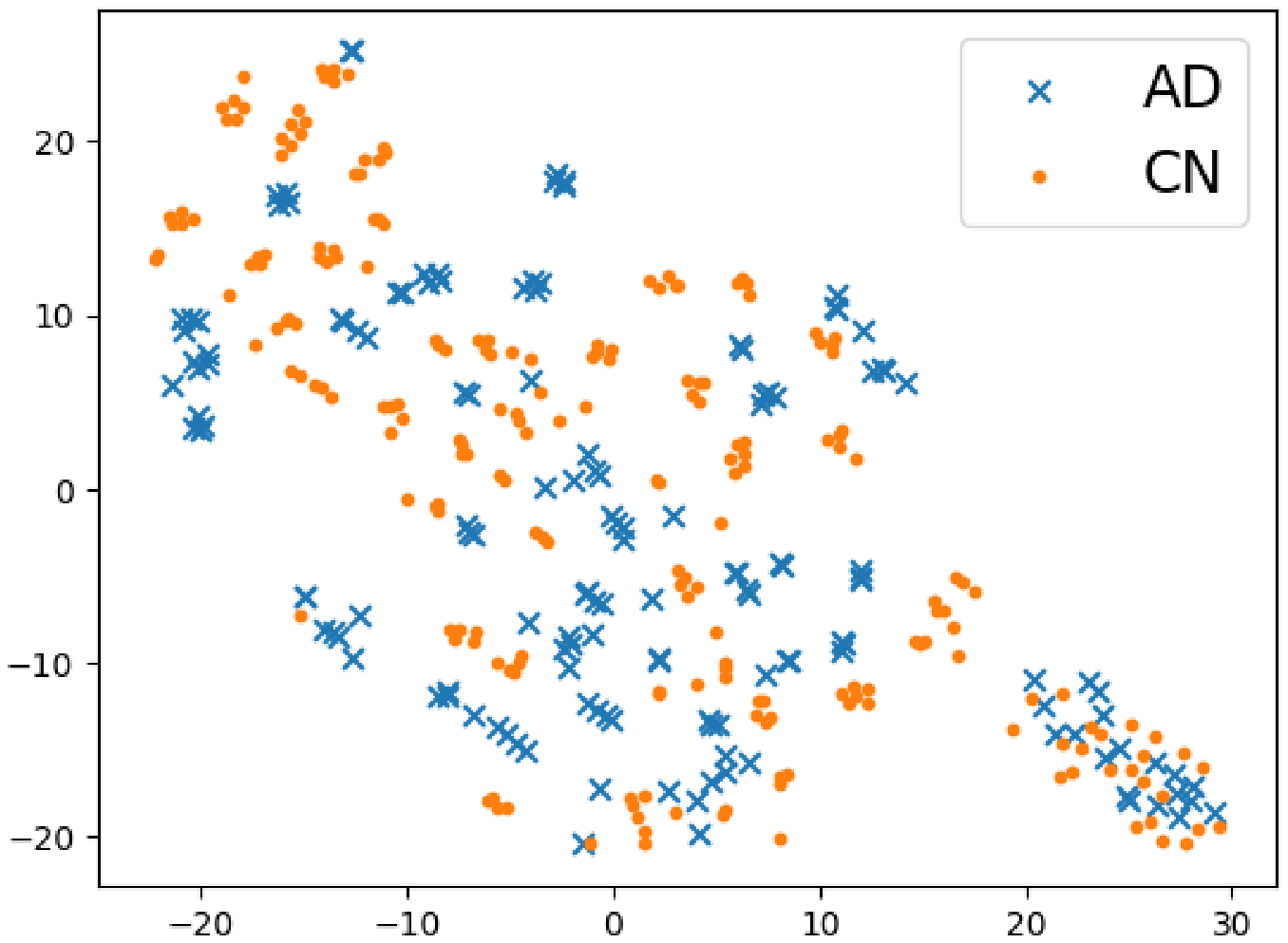}}
    \subfigure[DDCML]{\includegraphics[width=0.49\linewidth]{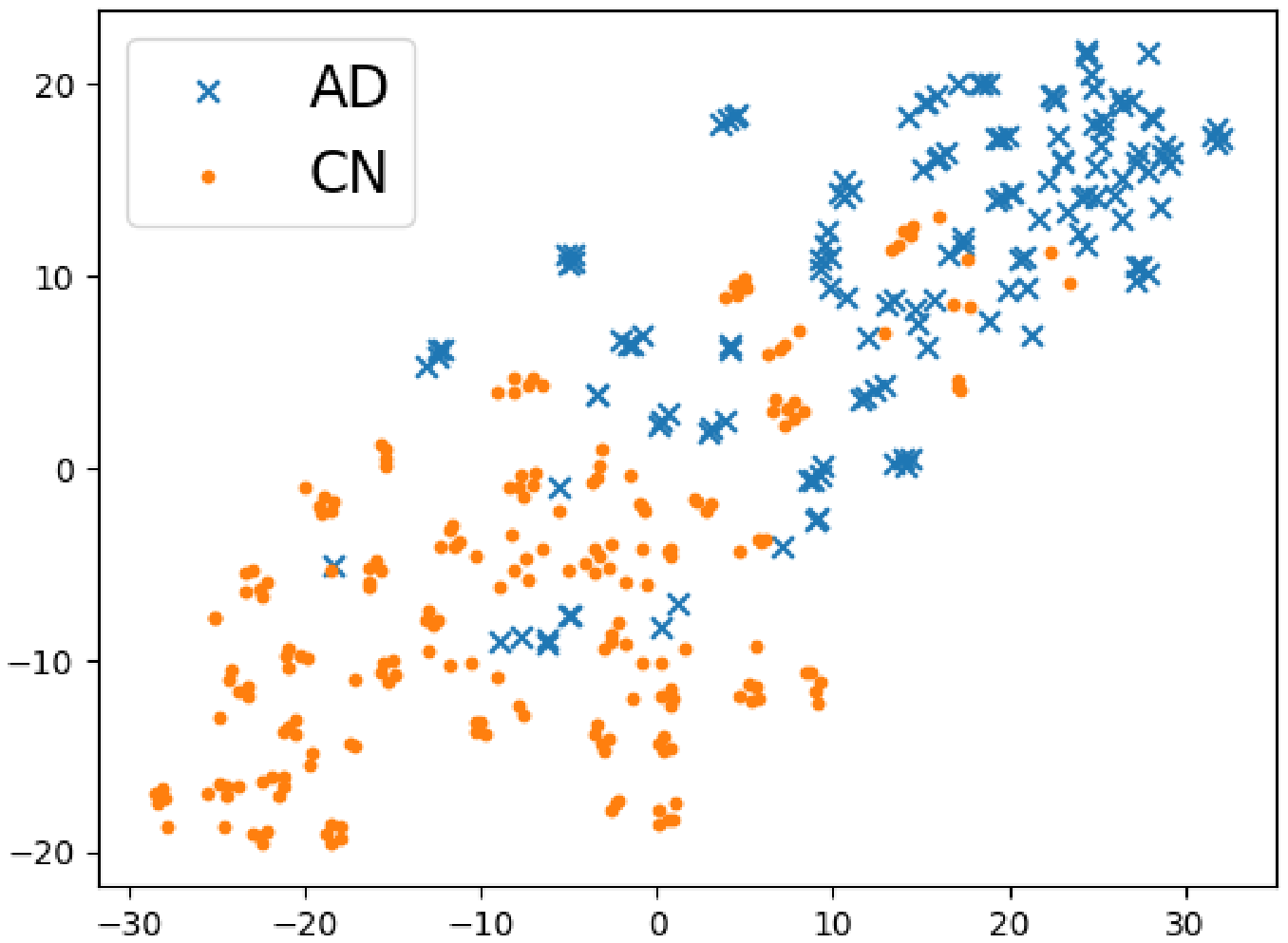}}
    \caption{Distribution of low-dimensional representations of AD and CN.}
    \label{fig:distribution}
  \end{figure}

  \figref{fig:distribution} compares the distribution of the low-dimensional representations obtained with plain 3D-CAE and the proposed DDCML scheme, using t-SNE~\cite{maaten2008visualizing}. Note that the presence of the intensity normalization described in Section \ref{subsec:I-norm} did not make a significant difference to this visual result and was omitted to save space.

\begin{figure}[]
    \subfigure[plain CAE]{\includegraphics[width=0.95\linewidth]{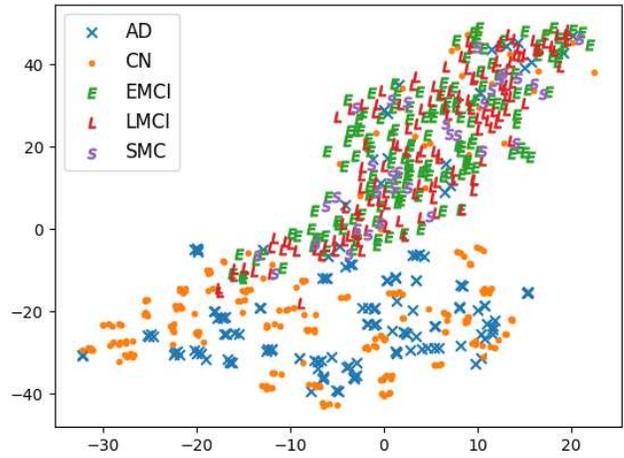}}
    \subfigure[DDCML]{\includegraphics[width=0.95\linewidth]{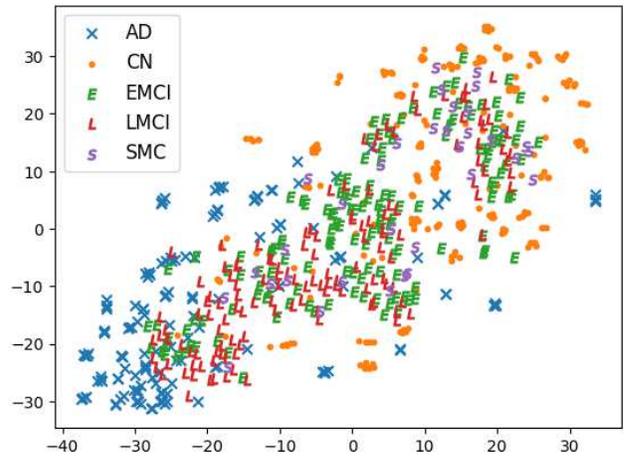}}
    \caption{Distribution of low-dimensional representations of all cases.}
    \label{fig:distribution_pizza}
  \end{figure}

  \figref{fig:distribution_pizza} shows the AD and CN cases overlaid with the distribution of LMCI, EMCI, and SMC data that was not included in the training. Note here that \figref{fig:distribution} and \ref{fig:distribution_pizza} show only one of five folds.

  In order to evaluate the effectiveness of our DDCML scheme, \tabref{tab:distances} summarizes the distance between the centroids of the data in each class in the low-dimensional feature space, with a normalized unit distance between AD and CN. The numbers in the upper-right diagonal are the distances in the proposed DDCML method, while the others are for plain 3D-CAE.

  \begin{table}[]
  \caption{Distance among cluster centroid}
  \begin{center}
  \begin{tabular}{lc|rrrrr}
  \hline
                       & \multicolumn{1}{l|}{} & \multicolumn{1}{l}{}   & \multicolumn{3}{c}{DDCML}                                                     & \multicolumn{1}{l}{}   \\
  \multicolumn{1}{c}{} &                       & \multicolumn{1}{c}{CN} & \multicolumn{1}{c}{SMC} & \multicolumn{1}{c}{EMCI} & \multicolumn{1}{c}{LMCI} & \multicolumn{1}{c}{AD} \\ \hline
                       & CN                    &                        & 0.311                   & 0.453                    & 0.655                    & 1                      \\
                       & SMC                   & 5.036                  &                         & 0.248                    & 0.474                    & 0.915                  \\
  plain CAE            & EMCI                  & 5.230                  & 0.522                   &                          & 0.242                    & 0.688                  \\
                       & LMCI                  & 5.192                  & 0.627                   & 0.345                    &                          & 0.489                  \\
                       & AD                    & 1                      & 5.064                   & 5.212                    & 5.135                    &                        \\ \hline
  \end{tabular}
  \end{center}
  \label{tab:distances}
  \end{table}

\section{Discussion}
  \tabref{tab:err_and_accs} shows that the plain 3D-CAE approach demonstrated excellent image-reconstruction performance from a highly compressed 150-dimensional feature space, in the same way as in~\cite{arai2018significant}; however, the two-class segmentation performance using K-means was just over 50\%. This indicates that the two categories are not well separated in the low-dimensional representation, and their direct application to CBIR is therefore less effective. Introducing metric learning markedly improves the clustering performance and is a key element in realizing CBIR, while the intensity normalization mitigates the slight decrease in the image reconstruction ability caused by metric learning. Finally, the features acquired by DDCML significantly improve the clustering performance over that of plain 3D-CAE (+29.1\%), minimizing the reduction in the image reconstruction performance (i.e. 1.24\% in RMSE and 0.017 in SSIM), as shown in \figref{fig:reconstructs}.
  From \figref{fig:distribution}, we can see that there are large differences in the distributions of the acquired features between the plain 3D-CAE method and our proposed scheme. In the former, the data are distributed regardless of the disease, whereas the latter scheme distributes data based on the disease to a certain extent . This result is obtained by further compressed for visualization, and thus we cannot conclude with this alone, but the proposed DDCML appears to provide a low-dimensional representation suitable for CBIR.
  In \figref{fig:distribution_pizza}, although only AD and CN were used for training, the distribution of all cases (AD, LMCI, EMCI, SMC, and CN) in the low-dimensional representation generated by the proposed DDCML scheme seems to be approximately distributed according to the severity of the disease category. In particular, SMC is only a subjective symptom of memory decline, and Alzheimer's symptoms are not observed. Since these are distributed near the CN results, the low-dimensional representations compressed using metric learning are considered suitable for CBIR.
  \tabref{tab:distances} shows that the DDCML successfully reproduced the relationships based on the severity of the disease category (CN$\fallingdotseq$SMC$\leq$EMCI$\leq$LMCI$\leq$AD), even though the three intermediate cases were not included in the training. This property is important in achieving CBIR.

\section{Conclusion}
  In order to realize a practical CBIR system for 3D brain MRI images, we propose a scheme called the disease-oriented data concentration with metric learning (DDCML) framework. DDCML consists of two key elements, intensity normalization and 3D convolutional autoencoders with metric learning, and these complement each other. DDCML can achieve an extremely efficient dimensional compression rate (4,096:1) while retaining the characteristics of the disease. In the near future, we will investigate and verify our DDCML framework using a wider variety of disease cases.

\section*{Acknowledgement}
This research was supported in part by the Ministry of Education, Science, Sports and Culture, Grant-in-Aid for Fun-damental Research (C), 17K08033, 2017–2020.

\bibliography{references}
\bibliographystyle{IEEEtran}

\end{document}